\documentclass[manuscript]{acmart}
\AtBeginDocument{%
  }

\setcopyright{acmlicensed}
\copyrightyear{2026}
\acmYear{2026}
\acmDOI{XXXXXXX.XXXXXXX}
\acmConference[CHI '26]{CHI Conference on Human Factors in Computing Systems}{May 23--28, 2026}{Barcelona, Spain}
\acmISBN{978-1-4503-XXXX-X/2026/05}

\usepackage{listings}

\begin{document}

\title{Empowering Users in Graph Rule Mining via Large Language Models}

\author{Francesco Cambria}
\affiliation{%
  \institution{Politecnico di Milano}
  \city{Milano}
  \country{Italy}}
\email{francescoluciano.cambria@polimi.it}

\author{Francesco Invernici}
\affiliation{%
  \institution{Politecnico di Milano}
  \city{Milano}
  \country{Italy}}
\email{francesco.invernici@polimi.it}


\author{Andrea Colombo}
\affiliation{%
  \institution{Politecnico di Milano}
  \city{Milano}
  \country{Italy}}
\email{andrea.colombo@polimi.it}

\author{Anna Bernasconi}
\affiliation{%
  \institution{Politecnico di Milano}
  \city{Milano}
  \country{Italy}}
\email{anna.bernasconi@polimi.it}

\renewcommand{\shortauthors}{Cambria et al.}


\begin{abstract}
In the era of interconnected data, graphs have emerged as an effective abstraction for modeling complex systems in an intuitive format, especially with the rise of Property Graphs, which offer an intuitive and scalable way of navigating non-intuitive structures. In this context, graph mining techniques have been developed for testing complex graph-based rules, as the {\tt MINE GRAPH RULE} operator, which, however, require users to have prior expertise both in graph theory and formal query language. In this work, we propose to bridge the gap between users and the graph-association rule-mining process by showing how Large Language Models (LLMs) can be easily prompted to formulate, refine, and interpret complex relational rules, directly producing {\tt MINE GRAPH RULE} queries.
\end{abstract}



\maketitle

\section{Introduction}
In the era of interconnected data, graphs have emerged as a premier abstraction for modeling complex systems effectively. Unlike traditional tabular formats, graph structures preserve and leverage the rich context of connections inherent in domains such as social networks~\cite{almabdy2018comparative}, financial infrastructure~\cite{henderson2020using}, and biological ecosystems~\cite{yoon2017use}. By representing entities as nodes and their interactions as edges -each enriched with specific attributes-  property graphs capture the complexities of real-world phenomena in which the relationships between data points are as significant as the data points themselves. However, this expressive power introduces substantial challenges, including query scalability and the difficulty of navigating non-intuitive, high-dimensional structures.

To extract value from these complex structures, traditional data mining techniques have been adapted to the property graph domain. Among these, association rule mining remains highly effective for uncovering latent insights within structured data. When applied to property graphs, this process seeks to identify associations within frequent topological patterns. To this end, various specialized methods have been developed, including Graph Pattern Association Rules (GPAR)~\cite{fan2015association}, Graph Association Rules (GARs)~\cite{fan2022association}, and Path Association Rule Mining (PARM)~\cite{sasaki2024mining}, among others.

Building upon these foundations, Cambria et al. \cite{cambria2025mine} introduced {\tt MINE GRAPH RULE}, an operator that establishes a formal syntax for extracting association rules directly from property graph databases. To facilitate practical application, in the original work, this operator was implemented as a stored procedure within Neo4j, one of the most widely adopted property graph database engines. Using the operator forces to adhere to a strict formal structure, leading users to more precise configurations and consistent graph association rules. Furthermore, the operator also defines connections with more complex logic (such as counting edges or evaluating multi-hop paths), ultimately extracting more nuanced rules.

Yet, despite the availability of such powerful operators, the structural complexity of graphs remains a significant barrier to accessibility for researchers without a computational background. Domain experts in sociology or medicine often possess profound intuitive knowledge of their data, but they are frequently hindered by the steep learning curve of the specialized query languages required for graph-based discovery. Recognizing this challenge, recent efforts have begun to explore more flexible interfaces; for instance, Le et al. \cite{le2025graph} exploit Large Language Models (LLMs) to better integrate domain experts within the mining pipeline.
Following this trajectory, this paper proposes using LLMs as effective mediators to bridge the gap between users and the graph-association rule mining process. By abstracting the syntactical complexities of the {\tt MINE GRAPH RULE} operator, our approach enables researchers to leverage graph mining techniques both accurately and intuitively. Ultimately, this framework extends graph mining capabilities to a broader audience, allowing domain experts to formulate, refine, and interpret complex relational rules without requiring prior expertise in graph theory or formal query languages.


\section{Mining a Graph database with LLMs}
A common feature of state-of-the-art LLMs is their ability to adapt to complex tasks based on instructions provided in a system prompt. In this work, we propose using this feature to assist non-technical users in extracting graph association rules by leveraging the {\tt MINE GRAPH RULE} operator.

We present the system prompt for this task in Figure~\ref{fig:mgr-prompt}.
We assign the LLM a specific role, framing its expertise within the scope of mining patterns from property graph databases.  Specifically, we define the LLM as an expert in association rule mining (using the \texttt{MINE GRAPH RULE} approach) and property graph database structures, with a focus on Neo4j and the Cypher query language. The instructions direct the LLM to interpret the database schema from user input and develop rules based on specific user interests, choosing the anchors and the itemset patterns. Furthermore, we attach some examples (Figure~\ref{fig:mgr-ex}) that present the operator to the LLM prompt, thereby enabling deeper technical understanding via few-shot learning. Following the original paper's framework, we adopt a schema centered on purchase relationships between Person and Item nodes, with each relationship classified by a specific Category node.
As shown in Figure~\ref{fig:mgr-output}, an LLM can generate MINE GRAPH RULE operators for various inputs. This confirms that state-of-the-art LLMs can master the required syntax while tailoring their outputs to user interests. 
The effectiveness of the system prompt (Figure~\ref{fig:mgr-prompt}) in translating user intent into valid {\tt MINE GRAPH RULE} syntax
has been preliminarily tested using three distinct LLMs (Gemini 3 Pro, Claude Sonnet 4.5, GPT-5 mini).
On each model, we ran 15 times the test using simple user inputs (e.g., those in Figure~\ref{fig:mgr-output}). The resulting outputs were manually validated against the operator's specific logic and parameters, yielding an outcome of 13 correct translations with GPT and 15 with Gemini and Claude. 

\begin{figure}[ht]
    \begin{minipage}{0.60\textwidth}
        \begin{lstlisting}[caption={}, frame=single]
(*\underline{Role}*):  
You are the Graph Rules Architect, an expert in Association Rules 
Mining (ARM) and Neo4j/Cypher. Your specialty is applying the MINE
GRAPH RULE (MGR) operator to property graphs. You possess a deep 
understanding of graph topology, path expressions, and the statistical 
metrics (Support and Confidence) required to extract interesting
patterns from interconnected data.
(*\underline{Core Competencies}*): 
Schema Interpretation: You can take a raw description of nodes 
and relationships and immediately identify the "Anchors" and
"Itemsets" for the MINE GRAPH RULE operator.
Operator Alignment: You must follow the syntax and semantics defined
in the "MINE GRAPH RULE" research. 1) Identify the anchor nodes, i.e., 
the "transaction" equivalent in the graph (the GROUPING ON node),
which will be the subject of the rule. 2) Define Body (X) and Head (Y) 
as conjunctions of relationship chains starting from the anchor, where 
each node has both variable and label while edges do not have a related 
variable. In a pattern all edges must be directed from left to right.
3) Use Support and Confidence as defined based on the MINE GRAPH RULE 
paradigm. Support is (*$C(R_i)/C(A)$*), where (*$C(R_i)$*) is the count of anchors 
matched by both Body and Head, and (*$C(A)$*) is the total anchor count). 
Confidence is (*$C(R_i)/C(B_i)$*), where (*$C(B_i)$*) is the count of anchors 
matched by the Body). 4) You are proficient in using WHERE limited to 
simple predicates on nodes properties, IGNORE for hiding specific 
intermediate node variables, countRel (>=n after the edge type) for 
relationship thresholds, and anyRel ({1,n}) for variable-length paths.
(*\underline{Instructions for Interaction}*):
Analyze Schema: Upon receiving a schema, identify the suitable anchor 
nodes and the itemset patterns.
Query Generation: Translate requirements into the operator, following 
precisely the syntax defined in the attached examples.
(*\underline{Style Guidelines}*):
Be technical, precise, and direct. 
Use code blocks for all MINE GRAPH RULE queries.\end{lstlisting}

\caption{LLM system prompt for writing the {\tt MINE GRAPH RULE} operator} 
\label{fig:mgr-prompt}
    \end{minipage}
    \hfill 
    \begin{minipage}{0.38\textwidth}
        \begin{lstlisting}[caption={}, frame=single]
(*\underline{Schema for Examples}*): 
Nodes: Person (properties: "age" (int)), 
Item (properties: "price" (int)), Category
Edges: BUY (from: Person, to: Item),
OF (from: Item, to: Category)
(*\underline{Examples}*):        
§MINE GRAPH RULE§ Example1
§GROUPING ON§ (p:Person)
§DEFINING BODY AS§ (p)-[:BUY]->(X:Item)
         §HEAD AS§ (p)-[:BUY]->(Y:Item)
§EXTRACTING RULES WITH SUPPORT§ > 0.1 
               §AND CONFIDENCE§ > 0.1 
§MINE GRAPH RULE§ Example2
§GROUPING ON§ (p:Person)
§DEFINING BODY AS§ 1..3 (p)-[:BUY]->(X:Item)
         §HEAD AS§ (p)-[:BUY]->(Y:Item)
§EXTRACTING RULES WITH SUPPORT§ > 0.1 
               §AND CONFIDENCE§ > 0.1    
§MINE GRAPH RULE§ Example3
§GROUPING ON§ (p:Person)
§DEFINING BODY AS§ (p)-[:BUY]->(X:Item)
                -[:OF]->(CX:Category)
         §HEAD AS§ (p)-[:BUY]->(Y:Item)
                 -[:OF]>(CY:Category)
§IGNORE§ X,Y
§EXTRACTING RULES WITH SUPPORT§ > 0.1
               §AND CONFIDENCE§ > 0.1 
§MINE GRAPH RULE§ Example4
§GROUPING ON§ (p:Person) §WHERE§ p.age > 18
§DEFINING BODY AS§ (p)-[:BUY]->(X:Item)
         §HEAD AS§ (p)-[:BUY]->(Y:Item)
§WHERE§ X.price > 50 and Y.price < X.price
§EXTRACTING RULES WITH SUPPORT§ > 0.1 
               §AND CONFIDENCE§ > 0.1 \end{lstlisting}
\caption{{\tt MINE GRAPH RULE} operator examples} 
\label{fig:mgr-ex}
    \end{minipage}
\end{figure}

\begin{figure}[ht]
    \begin{minipage}{0.32\textwidth}
        \begin{lstlisting}[caption={}, frame=single, 
    frameround=ffff]
(*\textbf{User Input}*):  
My graph has nodes: Patient 
(properties: age (int)), Symptom, and 
Disease; relationships HAS_SYMPTOM 
(from Patient to Symptom), 
DIAGNOSED_WITH (from Patient to 
Symptom), and PRESENTS_WITH (from 
Disease to Symptom). Write an 
operator for extracting association
rules from this graph to better 
diagnose diseases for fragile 
patients (consider a minimum support
of 0.1 and confidence of 0.1).
\end{lstlisting}
\vspace{-1mm}
\begin{lstlisting}[caption={}, frame=single, 
    frameround=ffff]
(*\textbf{LLM-Suggested Operator}*):
§MINE GRAPH RULE§ FragilePatient
§GROUPING ON§ (p:Patient) 
§WHERE§ p.age > 70
§DEFINING BODY AS§ (p)-[:HAS_SYMPTOM]
                        ->(s:Symptom)
        §HEAD AS§ (p)-[:DIAGNOSED_WITH]
                        ->(d:Disease)
§EXTRACTING RULES WITH SUPPORT§ > 0.1 
              §AND CONFIDENCE§ > 0.1
\end{lstlisting}
    \end{minipage}
    \hfill
    \begin{minipage}{0.32\textwidth}
        \begin{lstlisting}[caption={}, frame=single, 
    frameround=ffff]
(*\textbf{User Input}*):  
I want to extract association rules 
from a social network graph with 
nodes User with properties age (int),
gender (str), and hobby (str). 
Between users edges FOLLOW, 
LIKES_PICTURE_OF, and 
TAGS_IN_COMMENT can be drawn.
I am interested in association rules
that might predict that a social 
network user could follow another
based on its actions in the website
with 0.5 support and 0.8
confidence.
\end{lstlisting}
\vspace{-1mm}
\begin{lstlisting}[caption={}, frame=single, 
    frameround=ffff]
(*\textbf{LLM-Suggested Operator}*):
§MINE GRAPH RULE§ PathDiscovery
§GROUPING ON§ (u1:User)
§DEFINING BODY AS§ (u1)-[:FOLLOW]
     ->(u2:User)-[:FOLLOW]->(u3:User)
    §HEAD AS§ (u1)-[:FOLLOW]->(u3:User)
§IGNORE§ u2
§EXTRACTING RULES WITH SUPPORT§ > 0.5 
               §AND CONFIDENCE§ > 0.8§\end{lstlisting}
    \end{minipage}
    \hfill 
    \begin{minipage}{0.32\textwidth}
        \begin{lstlisting}[caption={}, frame=single, 
    frameround=ffff]
(*\textbf{User Input}*):  
I want to recommend clients of a
streaming platform what movies they
might like, using association 
rules with support and confidence
of 0.5. 
The streaming platform is 
based on a graph with nodes Client,
Movie, Genre, Actor, Director. 
Clients can WATCH a Movie, and LIKE
a Movie, Actor, or Director. Movies
are OF a Genre. Actor and Directors
are PART_OF and HAVE_DIRECTED Movies.
\end{lstlisting}
\vspace{-1mm}
\begin{lstlisting}[caption={}, frame=single, 
    frameround=ffff]
(*\textbf{LLM-Suggested Operator}*):
§MINE GRAPH RULE§ Recommendation
§GROUPING ON§ (c:Client)
§DEFINING BODY AS§ (c)-[:LIKE]
        ->(a:Actor)-[:PART_OF]
                        ->(m:Movie)
       §HEAD AS§ (c)-[:LIKE]->(m:Movie)
§IGNORE§ a
§EXTRACTING RULES WITH SUPPORT§ > 0.5 
               §AND CONFIDENCE§ > 0.5§\end{lstlisting}
    \end{minipage}
\caption{LLM conversation to generate different {\tt MINE GRAPH RULE} operators for disease, social and movie network} 
\label{fig:mgr-output}
\end{figure}


\section{Interpreting Graph Association Rules with LLMs}
With LLMs, users can mine association rules from property graphs using generated operators within the Neo4j environment. However, the high connectivity of graph data often results in a large number of rules that the user must manually evaluate. While support and confidence are standard metrics, they frequently fall short of identifying the most meaningful results, as significant insights can be buried under a large number of obvious or redundant rules. Furthermore, the raw output format of these rules can be counterintuitive for researchers outside the computer science or property graph database communities.

LLMs are also useful to address these post-processing challenges. Through a secondary system prompt, the model can be instructed to interpret and translate the output files into a more accessible format. This interpretation can also be conceptual; for example, rules with a confidence value of $1$ (or near $1$) can be interpreted as functional graph constraints, which are useful for graph repair and error detection.
Additionally, the LLM can be trained to rank these rules using complex logic tailored to the user’s specific expertise and goals. This allows for a more personalized filtering process that goes beyond simple statistical thresholds. The system prompt designed for this interpretation and ranking task is presented in Figure~\ref{fig:last-prompt}.

\begin{figure}[t]
\centering
\begin{lstlisting}
(*\underline{Role}*):  
You are the Graph Rules Architect, an expert in Association Rules Mining and Neo4j. You read the rules extracted using 
the MINE GRAPH RULE operator and you possess a deep understanding of graph topology, path expressions, and the 
statistical metrics (Support and Confidence) required to interpret the mined graph patterns.
(*\underline{Objective}*): 
Your goal is to read the set of rules  produced by the MINE GRAPH RULE operator, translate and explain them, and, 
finally, rank them according to the user instructions.
(*\underline{Core Competencies}*): 
Rules Translation: You can take the rules formatted into a dictionary structure and translate them using a Cypher-like 
syntax to make them more readable for the user. For each rule, the dictionary contains the structure of both body and
head -following the structure defined by the MINE GRAPH RULE operator - and the values of support and confidence.
Rules Interpretation: You can understand the deeper meaning of a rule. You can interpret and explain both the semantics 
of the graph pattern as well as its implication given the values of support and confidence according to the traditional 
applications of association rules mining.
(*\underline{Instructions for Interaction}*): 
Analyze the request: Upon receiving a request for a criterion for specific ranking of the rules other than support and 
confidence, evaluate the best strategy on how to apply it to the results. In addition, account for any other conditions 
set by the user - for instance, if they want only a subset of the rules filter them accordingly
Rules processing: Apply the user request for re-evaluation of the rules on their own criteria.
Rules translation: Translate into a Cypher-like syntax the results. Generally, explain them and reflect on possible
applications focusing on the user requests.
(*\underline{Style Guidelines}*): 
Be technical, precise, and direct. Use code blocks for all Cypher-like syntax.
\end{lstlisting}
\vspace{-2mm}
\caption{LLM System prompt for interpreting graph association rules} 
\label{fig:last-prompt}
\vspace{-2mm}
\end{figure}


\section{Conclusion}
A significant challenge in graph association rule mining is making the process accessible to non-technical users. Although the extracted patterns are powerful, their utility is limited if they cannot be easily interpreted or applied by domain experts who lack specific technical backgrounds in property graph databases.
Carefully designed LLM prompting can bridge this gap. These models can simplify the use of extraction tools and help users navigate the vast space of generated rules.
By using the prompts we developed, users can interact with two specialized agents that bring graph association rule mining closer to the end user. When used together, these agents create a refinement loop: the first agent generates initial rules based on the user's general interests, while the second agent ranks those results based on the user's specific domain expertise. This ranking often sparks more refined ideas or highlights interesting patterns, allowing the user to return to the first agent to mine even more specific and valuable rules.




\bibliographystyle{ACM-Reference-Format}
\bibliography{bibliography}

@String{Computer = "{IEEE} Computer" }

@String{Springer = "Springer-Verlag" }

@article{yoon2017use,
  title={Use of graph database for the integration of heterogeneous biological data},
  author={Yoon, Byoung-Ha and others},
  journal={Genomics \& informatics},
  volume={15},
  number={1},
  pages={19},
  year={2017}
}

@inproceedings{almabdy2018comparative,
  title={Comparative analysis of relational and graph databases for social networks},
  author={Almabdy, Soad},
  booktitle={2018 1st International conference on computer applications \& information security (ICCAIS)},
  pages={1--4},
  year={2018},
  organization={IEEE}
}

@article{henderson2020using,
  title={Using graph databases to detect financial fraud},
  author={Henderson, Richard},
  journal={Computer Fraud \& Security},
  volume={2020},
  number={7},
  pages={6--10},
  year={2020},
  publisher={MA Business London}}

@article{fan2015association,
  title={{Association rules with graph patterns}},
  author={Fan, Wenfei and others},
  journal={Proceedings of the VLDB Endowment},
  volume={8},
  number={12},
  pages={1502--1513},
  year={2015},
  publisher={VLDB Endowment}
}

@article{fan2022association,
author = {Fan, Wenfei and others},
title = {{Discovering association rules from big graphs}},
year = {2022},
issue_date = {March 2022},
publisher = {VLDB Endowment},
volume = {15},
number = {7},
journal = {Proc. VLDB Endow.},
month = {mar},
pages = {1479–1492},
numpages = {14}
}

@inproceedings{sasaki2024mining,
  title={{Mining path association rules in large property graphs}},
  author={Sasaki, Yuya and Karras, Panagiotis},
  booktitle={Proceedings of the 33rd ACM CIKM},
  year={2024}
}

@article{cambria2025mine,
  title={{Mine Graph Rule: A new GQL operator for mining association Rules in property graph databases}},
  author={Cambria, Francesco and others},
  journal={The VLDB Journal},
  volume={34},
  number={4},
  pages={1--21},
  year={2025},
  publisher={Springer}
}

@inproceedings{le2025graph,
  title={{Graph consistency rule mining with LLMs: An exploratory study}},
  author={Le, Hoa Thi and others},
  year={2025},
booktitle={Proceedings of the 28th EDBT Conference 2025},
pages={748-754}
}


\end{document}